\documentclass[
  aps,
  prl,
  reprint,
  groupedaddress,
  nofootinbib,
  nobibnotes,
  floatfix
]{revtex4-2}

\usepackage{amsmath,amssymb,bm}
\usepackage{graphicx}
\usepackage{booktabs}
\usepackage{siunitx}
\usepackage[dvipsnames]{xcolor}
\usepackage[
  colorlinks=true,
  citecolor=blue,
  linkcolor=blue,
  urlcolor=blue
]{hyperref}

\begin{document}
\title{Beam-Driven Transverse Deflecting Structure for Femtosecond Electron-Beam Diagnostics}
\author{S. Tomin}
\email{sergey.tomin@desy.de}
\author{D. Bazyl}
\author{W. Decking}
\author{I. Zagorodnov}
\affiliation{Deutsches Elektronen-Synchrotron DESY, Notkestrasse 85, 22607 Hamburg, Germany}

\date{\today}

\begin{abstract}
High-resolution longitudinal phase-space (LPS) diagnostics are essential for X-ray free-electron lasers and advanced accelerators. Conventional radio-frequency transverse deflecting structures (TDSs) provide direct femtosecond-scale LPS measurements, but their substantial RF-power and infrastructure requirements strongly limit their deployment at multi-GeV beam energies. Here, we propose a beam-driven transverse deflecting structure in which a leading driver bunch, separated by one RF bucket from a trailing witness bunch under study, excites long-lived wakefields in a resonant cavity array.  By placing the witness bunch near a zero crossing of the wakefield, the bunch experiences an approximately linear time-dependent transverse kick. Electromagnetic simulations of the resonant structure, combined with start-to-end beam-dynamics simulations based on European XFEL parameters at a final beam energy of 14 GeV, demonstrate a temporal resolution of $\sim 1.6$ fs for a 500 pC driver bunch, with a clear scaling toward the sub-femtosecond regime at higher charge.
\end{abstract}
\maketitle

The generation and control of high-brightness relativistic electron beams is central to the operation of X-ray free-electron lasers (XFELs) and the development of advanced accelerator concepts \cite{Pellegrini2016, Ackermann2007, Emma2010, Esarey2009, EuXFEL, Albert2021}. In these machines, performance is determined by the six-dimensional phase space of the beam. The longitudinal phase space (LPS), characterized by the peak current, slice energy spread, energy chirp, and temporal structure, plays a key role in compression dynamics, collective effects, and FEL emission. Advancing photon science, including fresh-slice and multicolor FEL techniques \cite{Lutman2016, Wang2024}, attosecond pulse generation \cite{Marinelli2017, Duris2020,Malyzhenkov2020, Prat2023,Yan2024}, and seeding schemes \cite{Feldhaus1997, Amann2012, Inoue2019, Liu2023}, requires accurate measurement and control of the LPS with femtosecond resolution.

To access the LPS directly, radio-frequency (RF) transverse deflecting structures (TDSs) have become the standard tool. From the original LOLA design \cite{Altenmueller1964, Loew1965} to modern X-band systems \cite{Behrens2014} and variable-polarization PolariX structures \cite{Craievich2020}, these devices map longitudinal coordinates onto a transverse plane. However, high-resolution measurements require multi-megawatt RF power, precise phase control, and substantial installation length. As a result, TDS systems are limited in their deployment due to their complexity and infrastructure requirements. They are therefore typically installed only at a small number of dedicated diagnostic locations. 

Considerable effort has thus been devoted to alternative diagnostics. Indirect, non-invasive methods based on coherent radiation, such as transition or diffraction radiation \cite{Happek1991, Castellano2001}, Smith–Purcell radiation \cite{Blackmore2009}, and electro-optic sampling \cite{Wilke2002, Berden2004}, provide information on bunch length and current profiles. However, reconstructing the full LPS from such measurements is intrinsically limited. Passive wakefield streakers \cite{Bettoni2016, Seok2018, Dijkstal2022, Dijkstal2024, Tomin2023} offer an RF-free approach by using the self-induced wakefields of the beam to generate a time-dependent transverse kick. A key limitation of these devices is the nonlinear mapping between longitudinal position and transverse displacement, which requires model-dependent reconstruction and limits accuracy.

%In this Letter, we introduce a beam-driven resonant deflecting structure that combines the highly linear longitudinal-to-transverse mapping of active radio-frequency (RF) transverse deflecting structures (TDSs) with the simplicity of passive wakefield-based streakers.

In this Letter, we introduce a beam-driven resonant transverse deflector for direct longitudinal phase-space measurements without the need for externally powered RF systems. This approach combines the approximately linear temporal mapping of active RF deflectors with the hardware simplicity of passive wakefield devices.

\begin{figure}[t]
    \centering
    \includegraphics[width=\columnwidth]{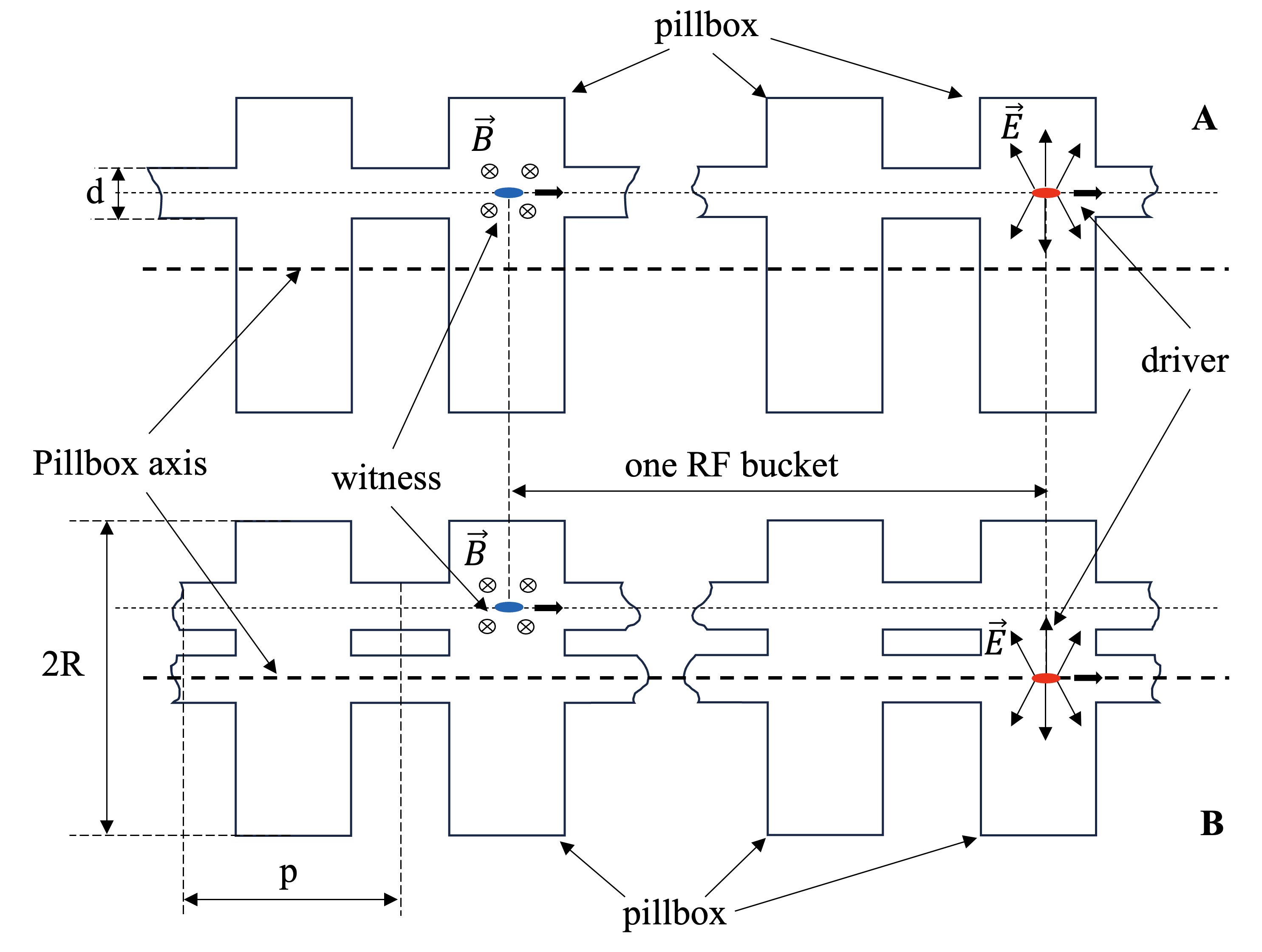}
    \caption{
    Schematic of the beam-driven resonant deflecting structure operating with an RF-bucket delay between the driver (red) and witness (blue) bunches. 
    (A) A purely collinear configuration, where both bunches propagate along the same trajectory, which is transversely offset from the central cavity axis. 
    (B) An alternative, dual-axis configuration, where the driver travels on the cavity symmetry axis to maximize wakefield excitation, and the witness bunch traverses an off-axis channel.
    }
    \label{img:pillbox}
\end{figure}

In conventional corrugated or dielectric-lined passive streakers, the relevant short-range wakefields decay rapidly because of open geometries and material losses, restricting the interaction to the fundamentally nonlinear self-streaking of a single bunch. In contrast, a leading ``driver'' bunch traversing a highly resonant cavity array, such as the sequence of pillbox cavities illustrated in Fig.~\ref{img:pillbox}A, excites long-lived transverse wakefields with quality factors exceeding \(10^3\). These oscillating wakefields persist long enough for a trailing ``witness'' bunch, delayed by one or more RF buckets, to experience a well-defined time-dependent transverse kick analogous to the streaking field of an active RF TDS.

To describe this interaction, consider a relativistic source bunch propagating through the structure with a transverse offset. The bunch excites electromagnetic fields in the surrounding cavity array, and a trailing witness electron experiences the corresponding Lorentz force. We use a right-handed coordinate system in which the \(z\)-axis is aligned with the nominal beam propagation direction, while \(x\) and \(y\) denote the horizontal and vertical transverse coordinates, respectively. The transverse position vectors of the source-bunch center of mass and the witness electron are denoted by \(\mathbf r_s=(x_s,y_s)\) and \(\mathbf r_w=(x_w,y_w)\), respectively, and their longitudinal positions by \(z_s\) and \(z_w\).

We define the three-dimensional vector wake potential as
\begin{equation}
\mathbf W(\mathbf r_w,\mathbf r_s,s)
=
\frac{1}{q_s}
\int
\left(
\mathbf E + \mathbf v_w \times \mathbf B
\right) dz ,
\end{equation}
where \(q_s\) is the source-bunch charge and the longitudinal separation is defined as \(s=z_s-z_w>0\), corresponding to a witness electron trailing the source bunch. Physically, \(\mathbf W(\mathbf r_w,\mathbf r_s,s)\) represents the integrated Lorentz force per unit source charge acting on the witness electron as it propagates through the structure. Its longitudinal component describes the energy change, while its transverse components describe the net transverse momentum kick.

As illustrated in Fig.~\ref{img:pillbox}, two interaction geometries are possible. In the collinear scheme, shown in Fig.~\ref{img:pillbox}A, the driver and witness bunches follow the same trajectory, transversely offset from the cavity symmetry axis. In the dual-axis scheme, shown in Fig.~\ref{img:pillbox}B, the driver travels on axis to enhance wakefield excitation, while the witness passes through a separate off-axis channel. Although this configuration may be more efficient, since it primarily excites TM-like monopole modes, it requires transverse separation of sub-nanosecond-spaced bunches, which is technically challenging. We therefore focus on the single-axis collinear geometry, Type~A, which avoids fast beam routing and can be implemented in a straight accelerator drift. The baseline structure used in this work has spatial period \(p=2.78\)~mm, beam-pipe diameter \(d=2.01\)~mm, and outer pillbox radius \(R=4.43\)~mm. The beam tube is shifted by \(1.86\)~mm from the pillbox axis, and the pillbox length is \(1.98\)~mm. These parameters were obtained from numerical optimization of the wakefield response at the witness-bunch position.

By timing the witness bunch close to the zero-crossing of the long-range transverse wakefield, the device generates an approximately linear longitudinal-to-transverse mapping, similar to an active TDS, and thereby reduces the need for model-dependent phase-space reconstruction. The required two-bunch pattern can be generated using either two synchronized photocathode laser pulses or a dedicated laser system capable of producing electron bunches in different RF buckets. Since the deflecting field is excited by the driver bunch, the structure does not require a local high-power RF system or associated water-cooling infrastructure. Apart from suitable optics and a downstream screen, it can therefore be integrated in locations where a conventional RF TDS would be impractical.

To quantify the performance of the proposed concept, we evaluated the electromagnetic response of the resonant structure using numerical simulations with \textsc{ECHO}~\cite{ECHO} and \textsc{CST Studio Suite}\textsuperscript{\textregistered}~\cite{CST}. Both time-domain wakefield simulations and eigenmode analyses were performed. In the simulations, the structure was modeled as copper with conductivity \(\sigma_{\mathrm{Cu}} = 5.8 \times 10^{7}\,\mathrm{S/m}\).

For given RF-bucket separation between the driver and witness bunches, the structure can be optimized either
through a modal coupling analysis or directly by maximizing the local wake
derivative from time-domain wake simulations, with the beam-tube offset from axis and
cavity parameters varied to place a strong transverse-wake zero crossing at the
desired witness delay.

\begin{figure}[t]
    \centering
    \includegraphics[width=\columnwidth]{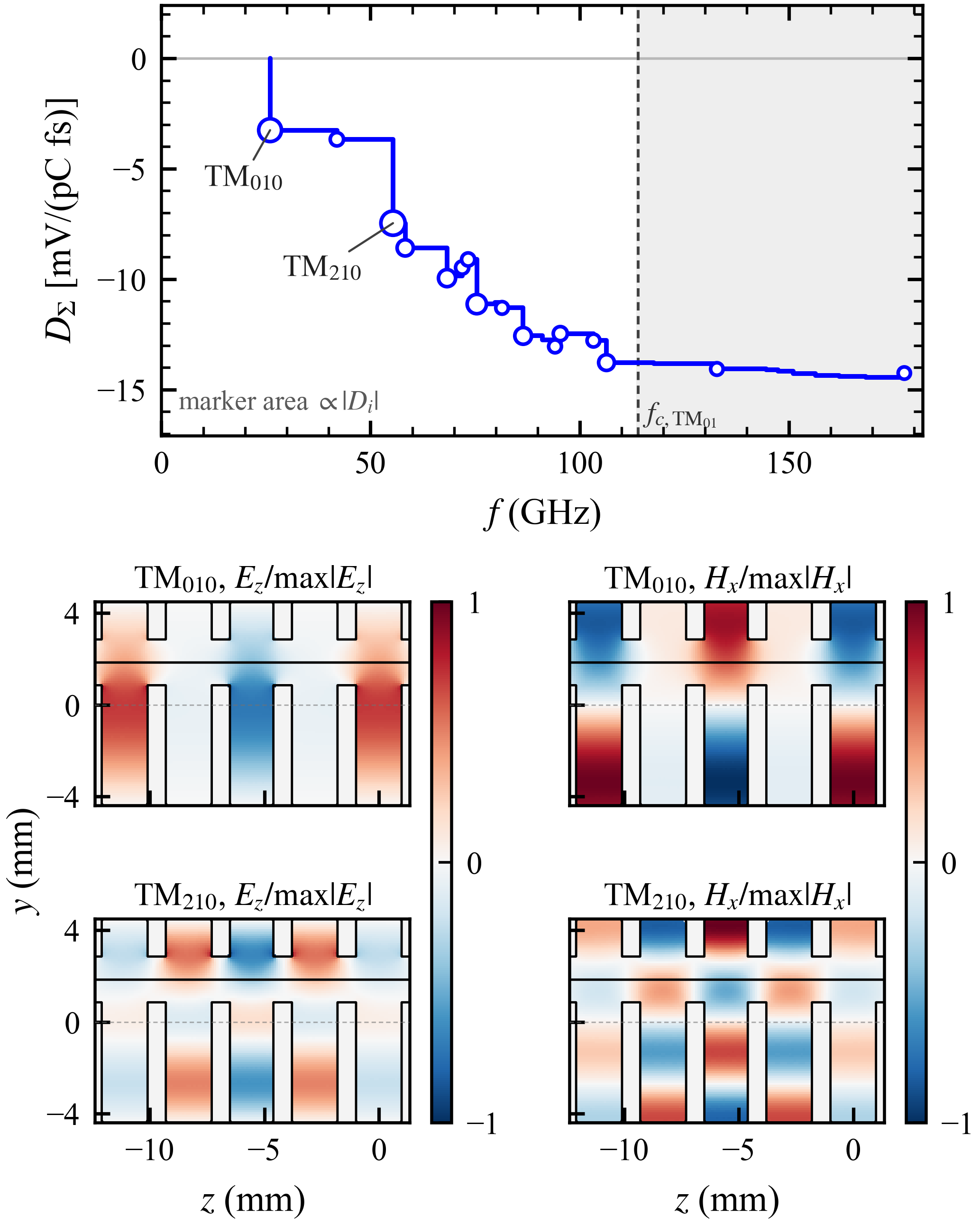}
\caption{
Electromagnetic mode content of the beam-driven resonant deflecting structure. The upper panel shows the cumulative modal contribution to the local transverse wake derivative at the witness-bunch position, expressed as the streaking strength per unit time. The marker areas are proportional to the absolute values of the individual modal contributions, \(|D_m|\). The shaded region indicates frequencies above the circular beam-pipe TM$_{01}$ cutoff, \(f_{c,\mathrm{TM}_{01}}\). The lower panels show the normalized field patterns of the two dominant contributing modes, TM$_{010}$-like and TM$_{210}$-like.
}
\label{img:rf_magic}
\end{figure}

The leading driver bunch excites a discrete spectrum of eigenmodes in the structure, as shown in Fig.~\ref{img:rf_magic}. For each eigenmode \(m\), the driver bunch couples to the longitudinal electric field sampled along the driver trajectory, while the witness bunch experiences the corresponding transverse Lorentz force along the witness trajectory. The modal contribution to the streaking field is determined by the product of the driver excitation and witness deflecting voltages, normalized by the stored modal energy. The local streaking strength at the witness position \(s_w\) is characterized by
\begin{equation}
D_m(s_w)
=
\left.\frac{dW_{y,m}}{dt}\right|_{s_w}
=
\omega_m A_{y,m}
\cos\!\left(
\frac{\omega_m s_w}{c}
+
\phi_m
\right),
\label{eq:modal-time-slope}
\end{equation}
where \(t=s/c\), \(c\) is the speed of light in vacuum, and \(A_{y,m}\) and \(\phi_m\) are the amplitude and phase of the modal wake potential, respectively. The cumulative contribution is obtained by summing all modes up to frequency \(f\),
\begin{equation}
D_\Sigma(f)=\sum_{f_m\le f}D_m .
\end{equation}

The upper panel in Fig.~\ref{img:rf_magic} shows this cumulative modal contribution evaluated at the witness position corresponding to a one-RF-bucket separation. A significant fraction of the total streaking strength is already provided by the first two modes, the TM$_{010}$-like and TM$_{210}$-like modes, which correspond, respectively, to the monopole TM$_{010}$ mode and the quadrupole TM$_{210}$ mode of an ideal closed pillbox cavity. Higher-order modes provide additional contributions, but the dominant physical mechanism remains the  deflection of the witness bunch by the transverse magnetic field component of the wakefield excited by the driver.

The TM$_{010}$-like and TM$_{210}$-like modes, which provide the largest contributions to the local wake slope, \(dW_y/dt\), at the witness-bunch position \(s_w\), are shown explicitly in the lower panels in terms of their normalized longitudinal electric-field and transverse magnetic-field distributions. These modes possess only a weak transverse electric field, and the transverse deflecting force is therefore dominated by the associated magnetic field. This is illustrated by the field maps in Fig.~\ref{img:rf_magic}, where the witness trajectory samples a region with a strong and approximately linear transverse magnetic field.

We note that active transverse deflecting structures typically operate in a dipole deflecting mode, most commonly a TM$_{110}$-like mode. In contrast, for the beam-driven structures considered in Fig.~\ref{img:pillbox}, the monopole TM$_{010}$-like mode provides a substantial contribution to the streaking field.

 For the copper structure, the dominant eigenmodes contributing to the transverse wake potential at the witness location have perturbative wall-loss quality factors of \(Q=(3.3\text{--}5.9)\times10^{3}\), corresponding to amplitude decay times \(\tau_A=Q/(\pi f)=\SIrange{13}{41}{ns}\). Thus, the fields remain essentially unchanged over the sub-nanosecond driver--witness delay, but decay strongly on hundred-nanosecond time scales and by many orders of magnitude on microsecond or longer time scales. At the European XFEL macropulse repetition rate of \(\SI{10}{Hz}\), the residual field from a previous driver-bunch shot is therefore negligible. Even for the \(\SI{4.5}{MHz}\) intra-train bunch spacing, corresponding to \(T_b\simeq\SI{222}{ns}\), the worst-case residual amplitude is bounded by \(\exp(-T_b/\tau_A)\lesssim4.5\times10^{-3}\) for the nominal \(Q\) values.

 Throughout this work, we use the European XFEL \cite{EuXFEL} as a representative example, assuming a bunch spacing of 769~ps corresponding to its 1.3~GHz RF frequency. The EuXFEL also provides a particularly demanding test case due to its beam energies of up to 17.5~GeV. 

\begin{figure}[t]
    \centering
    \includegraphics[width=\columnwidth]{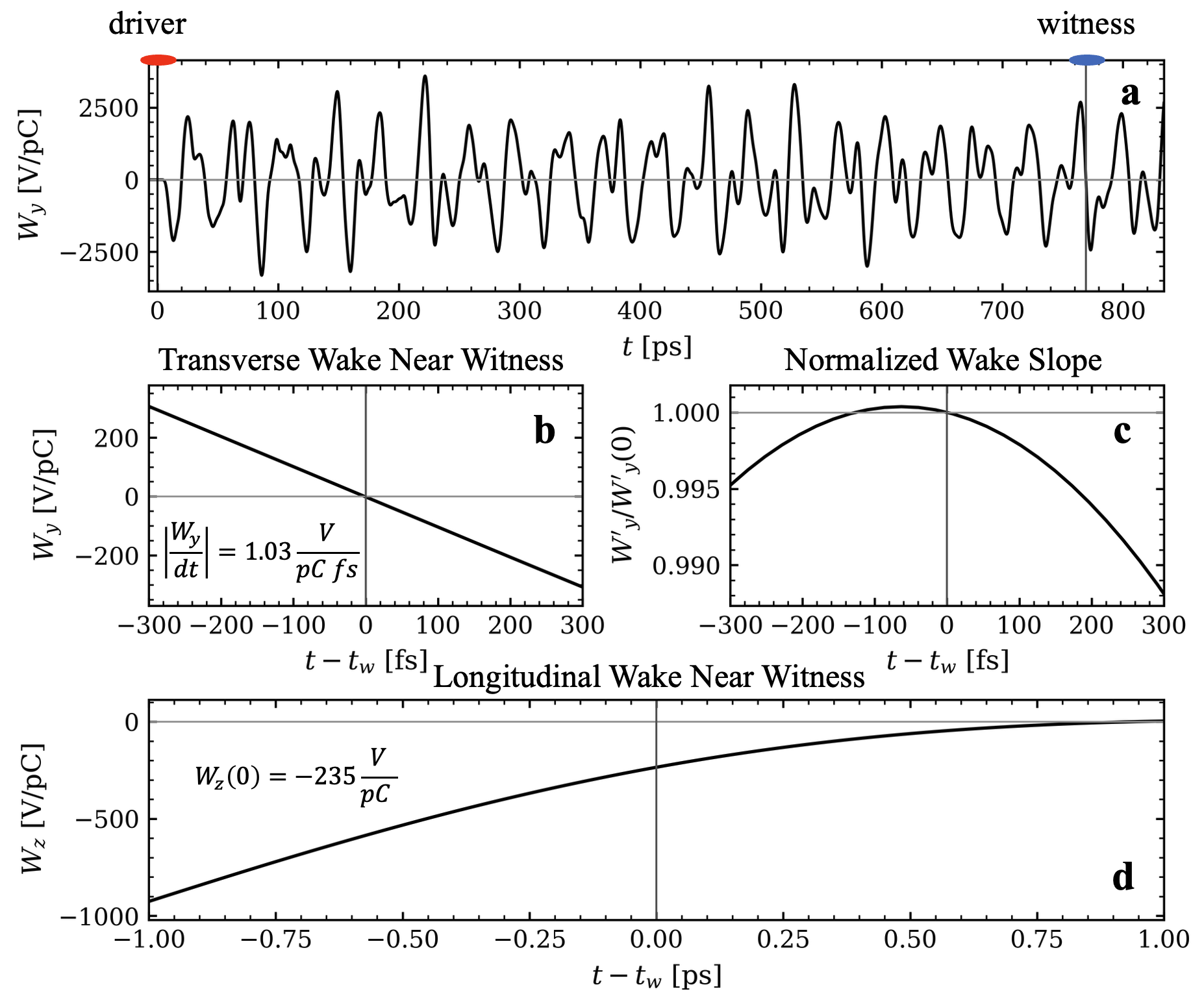}
    \caption{
Wakefields generated in a 1-m-long beam-driven transverse deflecting structure by a pencil bunch with a Gaussian longitudinal profile and an rms length of 0.5~mm.
(a) Long-range transverse wake potential \(W_y\) over the 769~ps separation between the driver and witness bunches.
(b) Local transverse wake near the witness bunch.
(c) Normalized transverse wake slope.
(d) Local longitudinal wake \(W_z\) near the witness bunch.}
    \label{img:wakes}
\end{figure}

The long-range wakefields induced by the driver bunch are shown in Fig.~\ref{img:wakes}. The resonant structure sustains an oscillating wake across the 769~ps delay. In the vicinity of the trailing witness bunch (Fig.~\ref{img:wakes}(b)), the local transverse wake slope reaches approximately $W'_{y} \equiv dW_{y}/dt \approx 1~\mathrm{V/(pC \cdot fs)}$ for a 1-m-long structure. As shown in Fig.~\ref{img:wakes}(c), the temporal streaking field remains linear, with the wake slope varying by 1\% over a $\pm 300$~fs interval. Concurrently, the long-range longitudinal wakefield, shown in Fig.~\ref{img:wakes}(d), decelerates the beam by 235~V/pC per meter of structure. For a 500~pC driver bunch, this corresponds to an average energy loss of 0.12~MeV per meter. At the multi-GeV beam energies typical of XFELs, this energy loss is dynamically negligible and does not significantly perturb the longitudinal phase space.

Because the induced wakefield in this regime closely reproduces the linear sweep of an active RF deflector, the resolving power of the beam-driven setup can be estimated analytically using standard TDS formalism \cite{Ding2011}. The temporal resolution is determined by
the ratio of the intrinsic unstreaked beam size at the observation screen, $\sigma_{y,0}=\sqrt{\varepsilon_y\beta_{y,\mathrm{scr}}}$, to the streaking factor
\begin{equation}
S =
\sqrt{\beta_{y,\mathrm{str}}\beta_{y,\mathrm{scr}}}
\sin(\Delta\psi_y)
\frac{q_d L}{E/e}
W'_{y},
\label{eq:S}
\end{equation}
yielding an RMS temporal resolution of $\sigma_t \simeq \sigma_{y,0}/|S|$. Here, $\varepsilon_y$ is the geometric emittance, $\beta_{y,\mathrm{str}}$ and $\beta_{y,\mathrm{scr}}$ are the vertical beta functions at the beam-driven structure and downstream screen, $\Delta\psi_y$ is the betatron phase advance between them (ideally $\pi/2$), $q_{\mathrm{d}}$ is the driver bunch charge, $L$ is the structure length, and $E/e$ is the beam energy expressed in volts.

In Eq.~(\ref{eq:S}), $q_d L W'_y$ is the direct analogue of the transverse-voltage slope $dV_\perp/dt$ in a conventional RF TDS. For an RF deflector operated at zero crossing, $dV_\perp/dt\simeq (2\pi c/\lambda)V_0$, so the streaking strength scales with RF frequency. In the beam-driven structure, this frequency dependence is already contained in the local wake slope $W'_y$. The large value of $W'_y$ arises from the high-frequency resonant content of the wakefield. As shown in Fig.~\ref{img:rf_magic} and discussed above, the lowest relevant mode is already near 26 GHz, more than twice the 12-GHz frequency of state-of-the-art X-band TDS systems, with higher-frequency modes further increasing the local wake slope.

To translate these wakefield properties into practical performance metrics, we evaluate the system using representative parameters of standard EuXFEL user operation at 14\,GeV. Furthermore, we assume a total structure length of $L = 6$\,m, comparable to the footprint of the proposed post-undulator active PolariX TDS installation, together with a typical normalized emittance of $\epsilon_{n,y} = 0.5\,\mu\mathrm{m\cdot rad}$ and a beta function at the beam-driven TDS location of $\beta_{y,\mathrm{str}} = 160$\,m. Substituting these values into Eq.~(\ref{eq:S}), the temporal resolution scaling becomes
\begin{equation}
\sigma_t \approx \frac{4.7~\mathrm{ps \cdot pC \cdot m}}{q_{\mathrm{d}}[\mathrm{pC}] \cdot L[\mathrm{m}]}.
\end{equation}

For a driver charge of 500\,pC, the estimated temporal resolution reaches approximately 1.6\,fs. Further optimization appears readily accessible. In particular, the European XFEL is designed to operate with bunch charges up to 1\,nC \cite{XFELTDR2006}, providing a direct operational pathway toward sub-femtosecond resolution through increased driver charge. In addition, the modular nature of the beam-driven structure allows straightforward extension of the interaction length, with the primary practical limitation arising from mechanical alignment tolerances. These estimates indicate that a beam-driven diagnostic could achieve temporal resolution competitive with, and potentially exceeding, that of state-of-the-art X-band RF transverse deflecting systems.

A practical challenge of the beam driven TDS is the separation of the driver and witness bunches on the downstream observation screen. When the witness bunch is positioned exactly at the zero-crossing of the transverse wakefield, it experiences no net dipole kick. Consequently, the unstreaked driver bunch and the streaked witness bunch would overlap transversely on the screen.

To avoid this ambiguity, a small timing offset can be introduced into the two-bunch spacing. This slight detuning from the exact zero-crossing imparts a constant dipole kick to the witness bunch while preserving the local linearity of the streaking field. The resulting transverse centroid shift spatially separates the projected longitudinal phase space of the witness bunch from the driver bunch on the observation screen.

To validate the measurement concept and verify the bunch separation, start-to-end beam tracking simulations were performed for a 500\,pC driver beam and 250\,pC witness beam using \textsc{Ocelot} \cite{Agapov2014,TominIPAC17}. The model of the beam-driven TDS includes the long-range transverse and longitudinal wakefields generated by the driver bunch, together with short-range longitudinal self-wake effects calculated using the formalism of Ref.~\cite{Bane2007}. Short-range transverse wakefields remain negligible due to the symmetric on-axis beam propagation. To spatially separate the witness bunch from the driver on the downstream observation screen, the witness bunch timing was detuned by 50\,fs from the exact transverse wakefield zero-crossing, thereby introducing a small net dipole kick while preserving the local linearity of the streaking field.

\begin{figure*}[t]
    \centering
    \includegraphics[width=\textwidth]{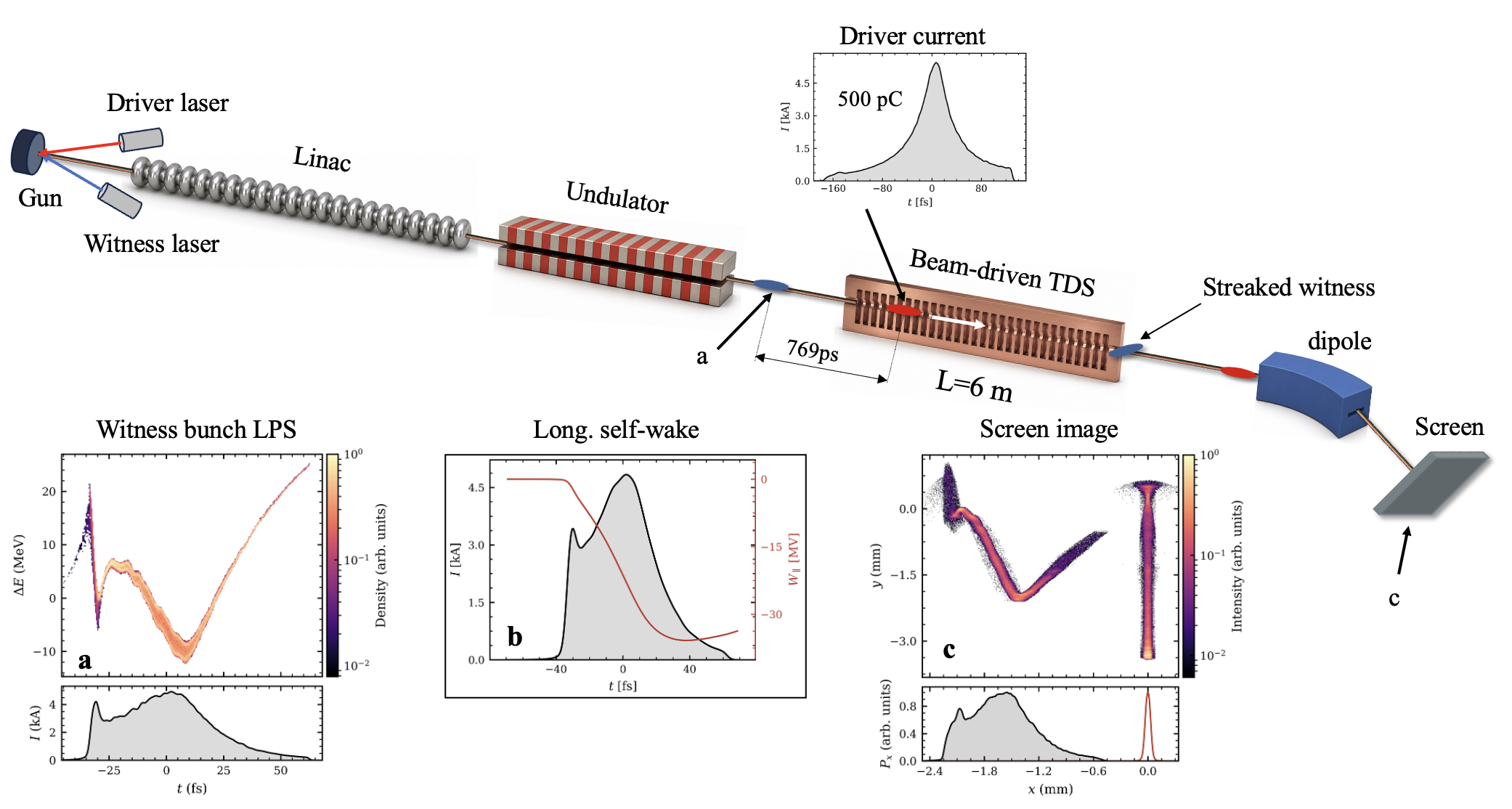}
    \caption{
    Schematic layout and start-to-end tracking simulations of the beam-driven transverse deflecting diagnostic. A leading driver bunch excites wakefields inside a 6\,m-long passive resonant structure, while a delayed witness bunch experiences transverse streaking. The two bunches are separated by one RF bucket ($769$\,ps). 
    (a) Longitudinal phase space and current profile of the witness bunch upstream of the beam-driven TDS. 
    (b) Local short-range longitudinal wake potential in the vicinity of the witness bunch together with the corresponding current profile. 
	(c) Simulated transverse distributions of the driver and witness bunches
on the downstream observation screen after transport through the dipole
spectrometer.
    }
    \label{img:tracking}
\end{figure*}

Tracking results for the proposed diagnostic beamline (Fig.~\ref{img:tracking}) show a clear spatial separation between the unstreaked driver and the streaked witness bunch on the observation screen. This isolation enables direct, single-shot measurement of the witness longitudinal phase space.

Due to short-range longitudinal self-wakefields, the witness bunch accumulates an average energy loss of approximately 21 MeV for a 250 pC bunch in a 6 m structure, together with a correlated intra-bunch energy variation visible in the tails of Fig.~4(b,c). However, because this deterministic self-wake acts strictly longitudinally and depends solely on the known static cavity geometry and bunch current profile, its effect can be analytically calculated and precisely deconvoluted from the final LPS reconstruction without degrading the temporal resolution.

Accurate time calibration is an essential component of any time-resolved beam diagnostic. In a conventional RF transverse deflecting structure (TDS), the calibration factor between the transverse screen coordinate and the longitudinal bunch coordinate is obtained by varying the RF phase and measuring the corresponding shift of the beam centroid on the observation screen. In the proposed beam-driven deflector, an analogous calibration can be performed by varying the relative delay between the driver and witness bunches and recording the corresponding transverse displacement of the witness bunch. The relative timing shift can be independently monitored using bunch arrival monitors (BAMs). At the European XFEL, the BAM system provides a single-shot timing resolution of approximately 3~fs rms and sub-femtosecond precision for averaged timing measurements \cite{Czwalinna2021}.

We evaluated the effect of long-range wakefields generated by a
\(500~\mathrm{pC}\) driver bunch on a \(250~\mathrm{pC}\) witness bunch in the
three-bunch-compressor layout of the European XFEL. The model includes
contributions from the \(1.3~\mathrm{GHz}\) cryomodules, the
\(3.9~\mathrm{GHz}\) third-harmonic cryomodule, the injector TDS, and the
accumulated \(1.3~\mathrm{GHz}\) linac wake contribution. For nominal
operating conditions, with the witness bunch compressed to a peak current of
approximately \(5~\mathrm{kA}\), the driver-induced long-range wakefields
perturb the final compression by \(15\text{--}25\%\), depending on the
compression settings, with the dominant contribution arising from the
third-harmonic cryomodule. Because the effect is determined primarily by the resonant structure
geometry and the fixed bucket spacing, it is highly reproducible and depends
only on the driver bunch charge. It can therefore be compensated through a
fixed RF retuning of the accelerator modules, independent of the witness
compression scenario. Analytical estimates and start-to-end tracking show
that the final bunch compression can thereby be restored to within better than
\(2\%\).

In conclusion, we have proposed a beam-driven transverse deflecting structure for direct femtosecond-scale longitudinal phase-space diagnostics. The concept combines the approximately linear streaking of active RF transverse deflectors with the simplicity of passive wakefield devices by using long-lived resonant wakefields excited by a preceding driver bunch. Electromagnetic and start-to-end simulations using European XFEL parameters demonstrate temporal resolution at the few-femtosecond level with a pathway toward the sub-femtosecond regime. The proposed approach provides a compact and RF-free alternative to conventional transverse deflecting systems for high-resolution diagnostics in advanced accelerator facilities. 

Beyond beam diagnostics, this beam-driven deflection principle offers a versatile tool for ultrafast beam manipulation. The highly linear time-dependent kick naturally lends itself to fresh-slice FEL techniques, where distinct temporal slices of a bunch are transversely separated to lase in different undulator sections. Furthermore, the structure could serve as a passive, high-repetition-rate kicker for multi-beamline bunch distribution in continuous-wave (CW) accelerator facilities. By exploiting the alternating polarity of the persistent oscillating wakefield, a single unperturbed driver bunch can cleanly bifurcate a beam train by sequentially deflecting trailing witness bunches into opposite transverse directions. This provides a zero-RF, infrastructure-light solution to some of the most pressing longitudinal and transverse beam-routing challenges in modern advanced accelerators.

\end{document}